# Excitation of Large-$k_\theta$ Ion-Bernstein Waves in Tokamaks

E. J. Valeo and N. J. Fisch

*Princeton Plasma Physics Laboratory, Princeton University, Princeton, NJ 08543*

The mode-converted ion-Bernstein wave excited in tokamaks is shown to exhibit certain very interesting behavior, including the attainment of very small poloidal phase velocities, the reversal of poloidal direction, and up-down asymmetries in propagation and damping. Because of these effects, this wave holds promise for channeling $\alpha$-particle power to ions, something that would make a tokamak fusion reactor far more attractive than presently envisioned.

One way to achieve significantly improved tokamak performance is to direct fusion $\alpha$-particle power into either electrons to sustain the plasma current [1] or into ions to increase the plasma reactivity [2]. Such a channeling of the $\alpha$-particle power could make possible the operation of fusion reactors in regimes in which there is a significant hot, nonmaxwellian component to the fuel ions, or in which the fuel ion temperature can be much greater than the electron temperature [3,4]. The fusion reactivity is then greatly enhanced, making possible a much more economical reactor.

The free energy in the $\alpha$-particles, namely, the free energy associated with the gradient in $\alpha$-particle pressure, may be tapped through a resonant quasilinear interaction with a poloidally propagating short wavelength electrostatic wave [1]. Such a wave diffuses the $\alpha$-particles both in space and energy, rather than just in energy. The free energy may be accessed if $k_\perp \rho_\alpha \gg 1$, where $\rho_\alpha \equiv v_\alpha/\Omega_\alpha$ is the $\alpha$-particle gyroradius, and where, for radial diffusion, $k_\perp \simeq k_\theta$. Upon interaction with the wave, the particle velocity is assumed displaced instantaneously in the $\hat{\theta}$ direciton by $\Delta v_\perp$, in the paralle direction by $\Delta v_\parallel$, and as a result the position of the guiding center changes by $\Delta \mathbf{v} \wedge \hat{\mathbf{i}}_\parallel/\Omega_\alpha$. (The confining magnetic field defines the $\parallel$ and $\perp$ directions; in cylindrical coordinates, $\hat{\theta}$ is the azimuthal or poloidal direction.)

Importantly, the change in the radial guiding center displacement, $\Delta r_{gc} = \Delta v_\perp/\Omega_\alpha$, is related directly to the energy change, $\Delta E = m_\alpha(v_\theta \Delta v_\perp + v_\parallel \Delta v_\parallel)$. This occurs because particles in a magnetic field are diffused by waves along the paths

$$v_\perp \Delta v_\perp + (v_\parallel - \omega/k_\parallel)\Delta v_\parallel = 0, \qquad (1)$$

and, since $k_\perp \rho_\alpha \gg 1$, an $\alpha$-particle is unmagnetized in the wave, and, hence, Landau resonant if at any point in its orbit

$$v_\theta = (\omega - k_\parallel v_\parallel)/k_\theta. \qquad (2)$$

It follows from Eqs. (1) – (2) that

$$\Delta r_{gc} = \Delta E k_\theta / m_\alpha \Omega_\alpha \omega. \qquad (3)$$

The usual resonance condition

$$v_\parallel = (\omega - n\Omega_\alpha)/k_\parallel. \qquad (4)$$

is recovered by requiring successive kicks add over many gyroperiods.

It is Eq. (3) that leads to the quest for large-$k_\theta$ waves. This is a necessary, if not sufficient, requirement for $\alpha$-channeling, and the focus of this paper. Clearly, in order to extract successfully $\alpha$-particle power, the key ratio $\Delta r_{gc}/\Delta E$ must neither be too large nor too small. In the former case, the $\alpha$-particle leaves the tokamak before giving up its energy. In the latter case, it is difficult to establish a population inversion. The ideal condition is for this ratio to be such that $\Delta r_{gc}$ is on the order of the minor radius for extracting the full $\alpha$-particle energy. From Eq. (4), this maximum excursion is

$$\Delta r_{\max} = (1/2)(k_\theta \rho_{\alpha 0})\rho_{\alpha 0}(\Omega_\alpha/\omega), \qquad (5)$$

where $\rho_{\alpha 0} \simeq 4\,\text{cm}$ is the initial gyroradius for 3.5 MeV $\alpha$-particles. For $1 < \Omega_\alpha/\omega < 3/2$, corresponding to wave frequencies between the deuterium and tritium resonances (such as the IBW), then for large $k_\theta$ sufficiently large, say $k_\theta \simeq 4\,\text{cm}^{-1}$, one obtains $32\,\text{cm} < \Delta r_{\max} < 48\,\text{cm}$, which is a range suitable for extraction, although toroidal effects can be expected to modify somewhat this range. The required diversion of power occurs if the wave then damps on fuel ions.

To achieve large $k_\theta$, one might consider the lower hybrid wave [1], but the wave may be difficult to excite in the proper regime, or there may be other difficulties [5], The mode-converted ion Bernstein wave [6], however, might be excited more easily with the requisite poloidal phase velocity and high wave number. This can occur as follows [7]: In a D-T (or other multispecies) plasma, the ion Bernstein wave (IBW) may be excited by launching a fast wave that mode-converts at the ion-ion resonance surface into an IBW [8]. This mode-converted ion Bernstein wave propagates essentially in the poloidal direction if two choices are made: one, the ion-ion hybrid resonance layer, which is more or less a vertical plane, is chosen to intersect near the tokamak magnetic axis; and, two, the fast wave is launched so as to intersect this vertical plane somewhat off the magnetic axis. Thus, horizontal is equivalent to poloidal propagation.

Of course, coupling to an IBW is possible through a variety of other methods [9]. One possibility would be to launch a slow electrostatic wave (electron plasma wave), which will subsequently convert to an IBW at the lower hybrid resonance. It would be difficult, however, in such a manner, to achieve the high poloidal wave numbers that are necessary, and that are achieved here, through mode conversion at a vertical ion-ion hybrid resonance. The method of coupling analyzed here, in contrast, exploits the production of the short wavelengths characteristic of a resonance layer that is aligned precisely perpendicular to the poloidal direction.

The location of this ion-ion hybrid resonance is determined by [10]

$$\omega^2 = (\Omega_D \Omega_T)\frac{f_D \Omega_T + f_T \Omega_D}{f_D \Omega_D + f_T \Omega_T}, \qquad (6)$$

where $\omega$ is the wave frequency, $\Omega_D \equiv eB/m_D$ is the deuterium gyrofrequency, $\Omega_T$ is the tritium gyrofrequency, and $f_D$ and $f_T$ are their respective fractional densities. The horizontal location of this resonance, which may be chosen so as to intersect the magnetic axis, may be controlled through the toroidal magnetic field, $B$, through the wave frequency, $\omega$, or through the relative concentration of the different ionic species.

Note that, since the gradient of the magnetic field strength in the major radius direction ($\hat{R}$) is the principal inhomogeneity intrinsic to the mode conversion process, the $\hat{R}$ component of the wavevector of the outgoing wave will increase most rapidly as the IBW propagates away from the mode conversion layer, thus achieving essentially poloidal propagation with high poloidal wavenumber.

In our analysis of the propagation of such a wave in a reactor-grade plasma, several things have been discovered: One, through a suitable choice of wave and tokamak parameters, and under conditions [4] important for channeling $\alpha$-particle power, namely ion temperatures about 20 KeV, and electron temperatures about 10 KeV, high $k_\theta$ can in fact be reached. Two, in this regime, the IBW can be made to reverse direction, so that the fields near high $k_\theta$ will be enhanced, as will the channeling effect. Three, it is possible to augment and exploit certain up-down asymmetries (because of the poloidal magnetic field) both for off-axis current-drive and for channeling $\alpha$-particle power. Four, a deuterium rich mixture achieves simultaneously both the high-$k_\theta$ channeling effect and the diverting of power to ions.

The dispersion relation, which describes the ion Bernstein wave, may be written in the form (see Eq. (1) of Ref. [9])

$$K_{xx} = -\frac{|K_{xy}|^2}{n^2} - n_\parallel^2 \frac{K_{zz}}{n_\perp^2 - K_{zz}}, \quad (7)$$

where the $K_{jj}$ are elements of the susceptibility tensor, and $\mathbf{n} \equiv \mathbf{k}c/\omega$ is the index of refraction. In the regime important for channeling, it is not possible [11] to employ conventional (small $k_\perp \rho$) approximations [12–14] of these susceptibilities. In what follows, the wave propagates horizontally ($\hat{x}$) and toroidally ($\hat{z}$), but not vertically ($k_y = 0$).

Because our interest is primarily in off-axis effects, the poloidal field importantly relates the parallel to the perpendicular wavenumber through $k_\parallel \simeq k_z + \theta k_x$, where $\theta \equiv B_\theta/B$. The first term, $k_z$, is the launched parallel wavenumber or the parallel offset. The second term represents the projection of the horizontal wavenumber, $k_x \simeq k_\theta$, on the magnetic field in the presence of a small poloidal magnetic field, $B_\theta$, which is assumed horizontal in the regime of interest. Suppose that the antenna launching the fast wave is up-down symmetric, and situated on the low-field periphery of the tokamak. Then one might envision two rays of fast waves propagating from the antenna, one that undergoes mode conversion to the IBW in the top half of the tokamak and the other in the bottom half. The resulting IBW spectrum, however, is not symmetric for nonzero poloidal field, since $B_\theta$ is up-down antisymmetric.

In Fig. 1, we trace such rays after the mode conversion for a 50:50 DT mixture. Note that for the case of cancellation of $k_\parallel$ ($\theta < 0$, or, say, for the IBW propagating in the top half of the tokamak), there is a tendency for the wave to reach sufficiently high $k_\perp$ to be of interest for channeling. Note also that, particularly for $k_z = 0.1$, the wave in fact turns around prior to damping, returning to the mode conversion layer at an upshifted wavenumber [15]. A caveat to the results here, and to be explored further, is that the ray tracing approach may not be valid, particularly for $\theta > 0$, where the damping is so intense as to occur in a distance short compared to a wavelength.

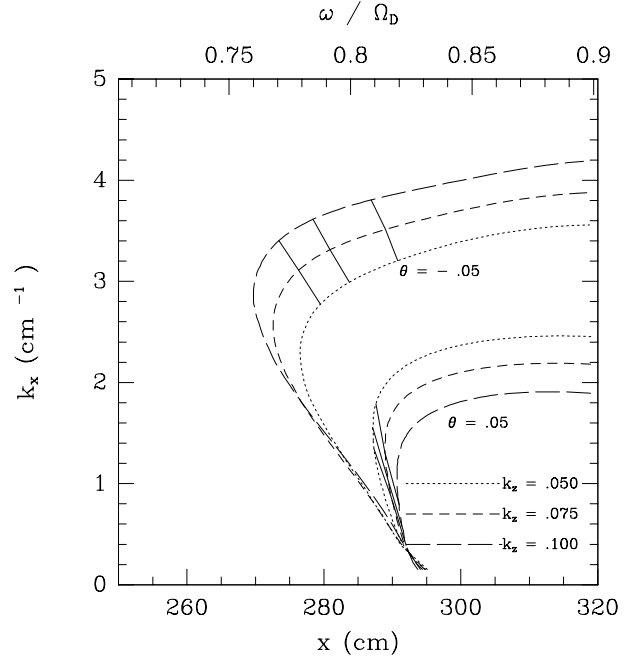

FIG. 1. Trajectory of IBW $k_x(x)$ from dispersion relation of Eq. (7), with parameters: $\omega = 1.69 \cdot 10^8 \text{ rad sec}^{-1}$, $n_e = 10^{14} \text{ cm}^{-3}$, $T_e = 10 \text{ keV}$, $T_D = T_T = 20 \text{ keV}$, with fractional densities, $f_D = f_T = 0.5$. The plasma inhomogeneity in the $\hat{x}$ direction is due entirely to the variation in magnetic field $\mathbf{B} = 4.8 \cdot 10^4 (\hat{z} + \theta \hat{x}) R/x$ gauss, with $\theta = \pm 0.05$, and "major radius" $R = 260$ cm. The broken curves are trajectories for the values .05, .75, and .1 cm$^{-1}$ of the toroidal wavevector $k_z$. The solid lines connect the coordinates at which .75, .5, and .2 of the initial power remains.

For channeling alpha power, it is significant both that the horizontal group velocity passes through zero and that this can be made to occur at high $k_\theta$. There is field enhancement where the group velocity vanishes, so that less power is required for the waves to diffuse $\alpha$-particles in a time short compared to their slowing down time.

Note also that there are considerable up-down asymmetries. First, clearly, the sense of poloidal propagation is opposite. Second, for $\theta > 0$, high-$k_\theta$ are not reached, since this wave damps immediately. Third, the damping for $\theta > 0$ is on ions, whereas, for $\theta < 0$, it is on electrons.

These asymmetries lead to different but not necessarily canceling effects. If the fast wave antenna is poloidally phased, then it would be possible to launch just the wave that produces the more desirable effect. However, this phasing is not easily achieved, and, in any event, present tokamaks are generally not equipped for such phasing. Hence, it is of interest, therefore, to examine whether, for example, extraction of $\alpha$-particle power can occur by the mode-converted IBW with toroidal phasing only.

For the extraction process, it is clear that the large disparity in $k_\theta$ leads to significant movement in the $\alpha$-particles from the top interaction with insignificant cancellation from the bottom one. In addition, there would be amplification of the wave by $\alpha$-particles at the top, but damping at the bottom. The only loss in not having poloidal phasing is that half the power is wasted with respect to extraction.

The current drive effect in the midplane [16], i.e. for $\theta = 0$, should not be affected by poloidal fields. The current drive effect off axis, however, is more complicated. Since the wave travels in opposite directions on top and bottom, and in both cases is completely damped, the current drive effect might be thought to vanish off axis, and, in fact, has been shown to vanish at least for very small $k_\parallel$ [17]. However, for realistic $k_\parallel$, there are interesting asymmetries: First, in the top case, the wave is 90% damped on electrons, whereas in the bottom case it is 2/3 damped on deuterium, 1/3 on tritium. Hence, the electrons and ions absorb momentum in opposite directions, but their currents from this effect *add*. Second, the parallel velocity of electrons and tritium ions that absorb wave power must be of the same sign as $\omega/k_\parallel$, whereas the parallel velocity of the resonant deuterium ions must be of opposite sign. Hence, to the extent that the resonant ions absorb power, there is the minority species heating current drive effect (MSHCD) that could be in opposite directions for these ions [18]. Depending on the effective bulk ion charge state, the trapped particle fraction, and other parameters, the MSHCD currents can actually be made to flow in either direction [19,20]. Thus, it is clear that sufficient asymmetry exists for the current drive effect even off axis, and that here the effects of the oppositely traveling waves can even be additive.

Note, however, that while extraction of $\alpha$-particle power appears to be in place, and while this power may in fact be channeled to produce current drive even with toroidal phasing only, it follows from Fig. 1 that this power can not be channeled to ions for increased reactivity. To accomplish this channeling effect, it will be necessary to explore further the parameter space. In Fig. 2, we vary the fuel mix, showing the damping on each species for (a) bottom case (i.e. $\theta > 0$), (b) $\theta = 0$, and (c) top case. (Note that the on-axis case has only 50% electron damping, which limits the mode-converted current drive, but the electrons for this case are relatively cold.)

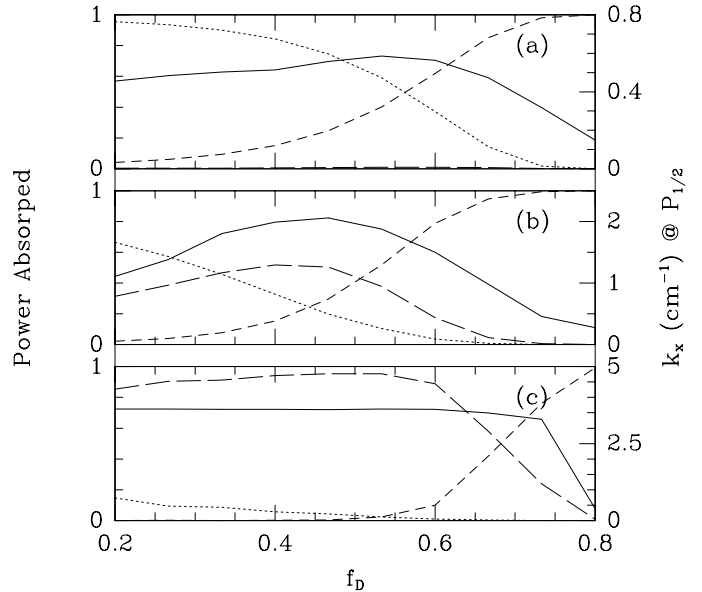

FIG. 2. Left axis displays fractional power absorbed by electrons (long dashes), deuterium (dots), and tritium (short dashes). Right axis displays $k_x$ for which 50% of power remains (solid line). All are plotted as a function of fractional deuterium density $f_D$, with $\theta = .05, 0, -.05$ in (a), (b), and (c), respectively.

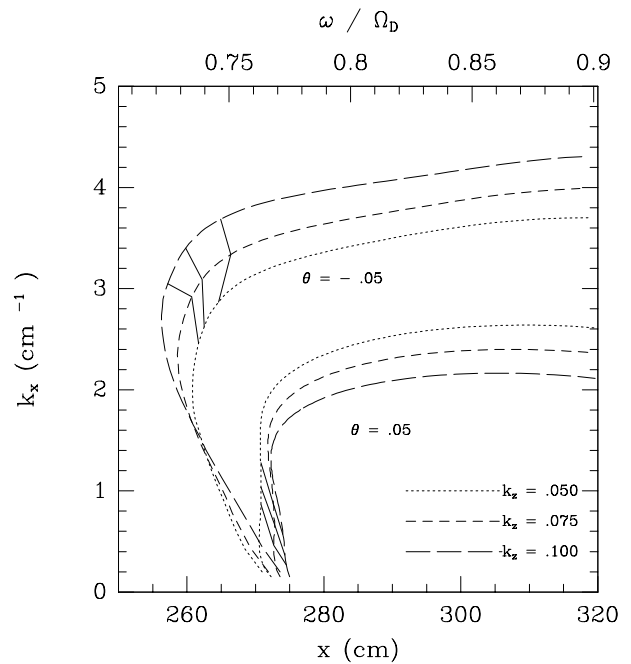

FIG. 3. Ray trajectories for $f_D = .7$. Other parameters are as in Fig. 1.

Of particular interest is that, for a 70:30 DT mix, there is the opportunity for both high $k_\theta$ and damping on tritium ions. Here, most of the power can be channeled to tritium ions at near maximum $k_\theta$. Note that this is quite relevant to reactor optimization, since at 70:30, there is 85% of the maxwellian fusion power, plus increased opportunity for an enhanced reactivity of a nonmaxwellian tritium tail [4].

In Fig. 3, we show the ray tracing for the 70:30 case, which indicates that here, as in the 50:50 case, even with toroidal phasing only, the channeling effect remains. However, the current drive effect on top and bottom largely cancels, since in both cases primarily tritium absorbs the power. As in the 50:50 case, the horizontal group velocity vanishes, and the dissipation, especially in the bottom trajectory, is over a very short distance. Also, note that the mode has shifted very much nearer to the tritium resonance.

Since the high-$k_\theta$ now occurs at only $\omega \simeq 0.74\Omega_\alpha$, it is of interest to check whether there is sufficient Doppler shift for $\alpha$-particles to interact with the wave. From Eq. (3), using $\rho_\alpha \simeq 4$ cm, and, from Fig. 3, taking $k_\perp \simeq 3\,\text{cm}^{-1}$ and $k_\parallel \simeq -.075\,\text{cm}^{-1}$, we find that $v_\parallel/v_\perp \simeq 1$, indicating that the resonance still falls within the bulk of the $\alpha$-particle distribution.

Although the effect reported is neither fully explored nor optimized here, it can be concluded from this study that large $k_\theta$ waves can be excited in tokamaks, off-axis, so as to divert $\alpha$-particle power to ions. At some sacrifice in optimization, both the diversion of power and current drive can be expected, even without poloidally asymmetric antenna phasing.

The channeling effect itself requires testing on tokamaks capable of D-T fusion, of which at present there are two, JET and TFTR. It would be important, however, first to observe the effects predicted here with respect to the up-down asymmetries in the magnitude of the wavevector, energetic particle loss, and current-drive. Mainly what is needed to observe these effects is off-axis excitation of the ion-Bernstion wave, with added benefit if this could be done asymmetrically either poloidally or toroidally.

The authors are indebted to C. F. F. Karney and R. Majeski for very useful comments. This work was supported by the United States Department of Energy under contract number DE–AC02–76–CHO3073.